\documentclass [11pt]{article}
\makeatletter
\def\section{\@startsection {section}{1}{\z@}{-3.5ex plus -1ex minus
 -.2ex}{2.3ex plus .2ex}{\large\bf}}
\def\subsection{\@startsection{subsection}{2}{\z@}{-3.25ex plus -1ex
minus -.2ex}{1.5ex plus .2ex}{\normalsize\bf}}
\makeatother
\makeatletter

\@addtoreset{equation}{section}

\makeatother
\newcommand{\be}{\begin{equation}}
\newcommand{\ee}{\end{equation}}
\newcommand{\bea}{\begin{eqnarray}}
\newcommand{\eea}{\end{eqnarray}}
\def\one{{\rm 1\kern -.9mm l}}
\begin{document}
\begin{titlepage}
\rightline{NORDITA-2002/17 HE} \rightline{DSF 5/2002}
%\rightline{\hfill November 2000}
\vskip 2.8cm
%\vskip 0.8cm
\centerline{\LARGE \bf More Anomalies from Fractional Branes
\footnote{Work partially supported by the European Community's Human
Potential Programme under contract HPRN-CT-2000-00131 Quantum
Spacetime and by MIUR under contract PRIN-2001025492.}}
\vskip 1.4cm
\centerline{\bf \large M. Bertolini $^a$, P. Di Vecchia $^a$,
M. Frau $^b$, A. Lerda $^{c,b}$,
R. Marotta $^{d}$}
\vskip .8cm \centerline{\sl $^a$ NORDITA, Blegdamsvej 17, DK-2100
Copenhagen \O, Denmark}
\vskip .2cm \centerline{\sl $^b$ Dipartimento di  Fisica Teorica,
Universit\`a di Torino} \centerline{\sl and I.N.F.N., Sezione di
Torino,  Via P. Giuria 1, I-10125 Torino, Italy}
\vskip .2cm \centerline{\sl $^c$ Dipartimento di Scienze e Tecnologie
Avanzate} \centerline{\sl Universit\`a del Piemonte Orientale, I-15100
Alessandria, Italy}
\vskip .2cm \centerline{\sl $^d$ Dipartimento di Scienze Fisiche,
Universit\`a di Napoli}
\centerline{\sl Complesso Universitario Monte S. Angelo, Via Cintia, I-80126
Napoli, Italy}
\vskip 1.8cm
\begin{abstract}
In this note we show how the anomalies of both pure and matter coupled 
${\cal N}=1,2$ supersymmetric gauge theories describing the low energy 
dynamics of fractional branes on orbifolds can be derived from 
supergravity.
\end{abstract}
\end{titlepage}
\renewcommand{\thefootnote}{\arabic{footnote}}
\setcounter{footnote}{0} \setcounter{page}{1}

%%%%%%%%%%%%%%%%%%%%%%%%%%%%%%%%%%%%%%%%%%%%%%%%%%%%%%%%%%%%%%%%%%%%%%
\tableofcontents
\vskip 1cm

%%%%%%%%%%%%%%%%%%%%%%%%%%%%%%%%%%%%%%%%%%%%%%%%%%%%%%%%%%%%%%%%%%%%
\section{Introduction}
Recently there has been a lot of activity in trying to generalize the
AdS/CFT correspondence and promote it to a more general gauge/gravity
duality for theories without conformal invariance or with less
supersymmetry. Indeed many different approaches have been proposed to
this aim, like for example the study of mass deformations of conformal
field theories and of the corresponding RG flows, or the study of
D-branes wrapping supersymmetric cycles of K3 and Calabi-Yau spaces,
or finally the study of fractional D-branes, in conifold and orbifold
backgrounds. 

Fractional branes in conifolds backgrounds were first considered in
Refs.\cite{KN,KT,KS} where a configuration of $N$ regular and $M$
fractional D3 branes on $T^{1,1}$ has been analyzed and
the corresponding supergravity solution has been derived. This
solution has been successfully used to describe the dual four
dimensional gauge theory, which has ${\cal N}=1$ supersymmetry, a
gauge group $SU(N+M) \times SU(N)$ and a non trivial matter content
\cite{KW}. While very interesting IR properties of this gauge theory
have been obtained by
considering the non-singular solution on a deformed 
conifold \cite{KS}, to describe the UV features of the gauge theory it is
enough to consider the singular solution of Ref.~\cite{KT} which
indeed accounts for the logarithmic running of the coupling constants
and the precise coefficients of the $\beta$-functions of the two gauge
groups \cite{HKO}. Very recently, in Ref.~\cite{KOW} it has been shown
that also the chiral anomaly and the breaking of the $U(1)$ R-symmetry 
to ${\bf Z}_{2M}$ are correctly incorporated in the classical 
UV supergravity solution of Ref.~\cite{KT}. 

Fractional D-branes in orbifold backgrounds have been extensively
considered in the recent literature. In Ref.~\cite{d3} the
supergravity solution of $M$ fractional D3 branes in a ${\bf C}^2/{\bf
Z}_2$ orbifold has been explicitly obtained (see also
Ref.~\cite{polch}) and used to derive the exact (perturbative)
$\beta$-function of the dual ${\cal{N}}=2$ $SU(M)$ Yang-Mills theory.
This analysis has been subsequently generalized by adding fractional
D7 branes that yield hypermultiplets in the gauge theory
\cite{grana2,d3d7}, or by considering D-branes in ${\bf Z}_k$
orbifolds \cite{marco} (see Ref.~\cite{REV} for a recent review on
fractional  branes in ${\cal N}=2$ orbifolds and a complete list of
references). More recently, similar results have been also obtained
for fractional D-branes in the ${\cal{N}}=1$ orbifold ${\bf C}^3/({\bf
Z}_2\times {\bf Z}_2)$ \cite{FERRO}.

In this paper, prepared while Ref.~\cite{KOW}, which contains also a
brief discussion of the chiral anomaly  for the ${\cal N}=2$ orbifold,
has appeared,  we show how the chiral anomalies of both ${\cal{N}}=2$
and ${\cal{N}}=1$ supersymmetric gauge theories in four dimensions can
be very simply obtained from the explicit supergravity solutions
describing fractional D3 branes in ${\bf C}^2/{\bf Z}_2$ and ${\bf
C}^3/({\bf Z}_2\times {\bf Z}_2)$ orbifolds presented, respectively,
in Refs.~\cite{d3,d3d7,REV} and Ref.~\cite{FERRO}.  In particular, we
show how to obtain the correct $\beta$-function and the chiral
anomaly of the pure ${\cal N}=1$ super Yang-Mills. It is worth to
emphasize that our present analysis does not rely on the probe
technique which instead was used in our previous papers. In fact, here
we will exploit a simple holographic representation of a gauge theory
operator in terms of supergravity bulk quantities and deduce from it
how the scale and chiral transformations are realized in the
supergravity description.

%%%%%%%%%%%%%%%%%%%%%%%%%%%%%%%%%%%%%%%%%%%%%%%%%%%%%%%%%%%%%%%%%%%%%
\section{${\cal{N}}=2$ gauge theories and fractional branes in ${\bf Z}_2$
orbifolds}

Let us consider a ${\cal{N}}=2$ super Yang-Mills theory in four
dimensions with gauge group $SU(M)$ and $N$ fundamental
hypermultiplets. As is well known (see for example
Ref.~\cite{HOUCHES98}), this theory has a $\beta$-function given by
\begin{equation}
\beta(g) = - \frac{(2M-N)}{16\pi^2}\,g^3 \label{beta}
\end{equation}
where $g$ is the running coupling constant, and a $U(1)_R$ anomaly
given by
\begin{equation}
\partial_{\alpha} J^{\alpha}_{R} = 2 ( 2M -N)\,\, q(x)~~~;~~~ q(x)=
\frac{1}{32 \pi^2} F^{a}_{\alpha \beta} {\widetilde{F}}^{\alpha\beta}_a
\label{ano78}
\end{equation}
where $J_R$ is the $R$-current and $q(x)$ is the topological charge
density. In writing these formulas we have used standard conventions,
namely we have normalized the generators of $SU(M)$ so that the
quadratic Casimir invariant is $M$ in the adjoint representation and
$1/2$ in the fundamental representation, and have assigned a
$R$-charge $1$ to the gauginos and $-1$ to the chiral fermions in the
hypermultiplets~\footnote{We note that in Ref.\cite{KOW} a different
$R$-charge assignment has been given to the various fields. In
particular their charges differ from ours by a factor of 1/3.}.

In supersymmetric theories the scale anomaly, proportional to the
$\beta$-function, and the chiral anomaly are in the same supersymmetry
multiplet. In the theory at hand, this fact can easily be seen by
introducing the complex quantity
\begin{equation}
\tau_{\rm YM} \equiv \frac{ \theta_{\rm YM}}{2\pi} + {\rm i}\, \frac{4
 \pi}{g^{2}}
\label{gau63}
\end{equation}
where $\theta_{\rm YM}$ is the Yang-Mills $\theta$-angle, and
observing that the term reproducing the anomalies has the following
form \cite{PN}
\begin{equation}
\tau_{\rm YM} = {\rm i} \,\frac{2M-N}{4 \pi} \log
\frac{\phi^2}{\Lambda^2} \label{tau69}
\end{equation}
where $\Lambda$ is the dynamically generated scale and $\phi$ is the
(complex) scalar field of the ${\cal N}=2$ vector multiplet.  As is
well known, due to ${\cal N}=2$ non-renormalization theorems,
Eq. (\ref{tau69}) is the complete perturbative result which is
corrected only by instanton effects. Then the scale and chiral
anomalies can be simply obtained by looking at the response of
$\tau_{\rm YM}$ to a rescaling of the energy by a factor of $\mu$ and
to a $U(1)_{R}$ transformation with parameter $\alpha \in [\,0,2\pi )$
respectively. Under these transformations the scalar field $\phi$,
which has the scale dimension of a mass and a $R$-charge 2 (remember
that the gauginos have $R$-charge 1), transforms as follows
\begin{equation}
\phi \rightarrow \mu\,{\rm e}^{2\,{\rm i}\, \alpha} \,\phi~~,
\label{tra34}
\end{equation}
and thus from Eq.(\ref{tau69}) we easily find that
\begin{equation}
\tau_{\rm YM} \rightarrow \tau_{\rm YM} + {\rm i} \,\frac{2M-N}{2\pi}
(\log\mu+2{\rm i}\alpha)~~. \label{chisca3}
\end{equation}
This equation implies that
\begin{equation}
\frac{1}{g^{2}} \rightarrow \frac{1}{g^{2}}
+\frac{2M-N}{8\pi^2}\,\log\mu~~~~~{\rm and}~~~~~\\ \theta_{\rm YM}
\rightarrow \theta_{\rm YM} -2\,(2M-N)\,\alpha~~, \label{tra39}
\end{equation}
which are equivalent to Eqs.(\ref{beta}) and (\ref{ano78})
respectively.

We now show how these results can be obtained in a very simple way by
using the supergravity solutions of fractional D-branes that we
presented in Refs.~\cite{d3,d3d7}. Let us recall that in string theory
the simplest way to realize a four dimensional ${\cal{N}}=2$
Yang-Mills theory with gauge group $SU(M)$ and $N$ hypermultiplets is
to consider a stack of $M$ fractional D3 branes and $N$ D7 branes of
type IIB in the orbifold ${\bf C}^2/{\bf Z}_2$ \cite{grana2,d3d7}. For
definiteness we assume that the ${\bf Z}_2$ parity acts as a
reflection in the directions 6789 (labeled by indices $\ell,m,...$),
and that the D3 branes extend along the directions 0123 (labeled by
indices $\alpha,\beta,...$) while the D7 branes wrap the directions
01236789. With this arrangement, the directions 4 and 5 are transverse
to both types of branes and define a plane where they can move. In
this plane it is convenient to use the complex coordinate
\begin{equation}
z\equiv x^4 +{\rm i}\,x^5 = \rho \,{\rm e}^{{\rm i} \,\theta}~~.
\label{zeta}
\end{equation}
The supergravity background created by this D3/D7 brane system
comprises the dilaton $\varphi$, the (Einstein frame) metric of the
form
\begin{equation}
ds^2 = H^{-1/2}\, \eta_{\alpha\beta}\,d x^\alpha dx^\beta + {\rm
e}^{-\varphi}\,H^{1/2} \left(d\rho^2+\rho^2d\theta^2\right) + H^{1/2}
\,\delta_{\ell m} dx^\ell dx^m ~~,
\label{metric}
\end{equation}
a R-R 0-form $C_{(0)}$, a R-R 4-form $C_{(4)}$, and two 2-forms,
namely $B_{(2)}$ from the NS-NS sector and $C_{(2)}$ from the R-R
sector, which are given by
\begin{equation}
C_{(2)} = c~ \omega_{(2)}~~~~~,~~~~B_{(2)} = b~ \omega_{(2)}
\label{wra98}
\end{equation}
where $\omega_{(2)}$ is the anti-self dual 2-form associated to the
vanishing 2-cycle of the orbifold ALE space. The explicit expressions
for the various fields have been derived in Ref.~\cite{d3d7} (see also
Ref.~\cite{grana2}), but for our present purposes it is enough to
recall that
\begin{eqnarray}
{\rm e}^\varphi &=& \frac{1}{1-\frac{N g_s}{2 \pi} \,\log
\frac{\rho}{\epsilon}} ~~,\label{dilsol} \\ C_{(0)}&=&
\frac{Ng_s}{2\pi}\,\theta ~~,\label{C0sol}\\ b &=& (2
\pi^2{\alpha'})\; \frac{1+\frac{(2M-N)g_s}{\pi} \, \log
\frac{\rho}{\epsilon}} {1-\frac{N g_s}{2 \pi} \,\log
\frac{\rho}{\epsilon}} ~~, \label{bsol}\\ &&\nonumber \\ c &=&
-(2\pi\alpha') \,g_s\,\left(2M - \frac{N}{2}\;\frac{1-\frac{2M
g_s}{\pi} \,\log \frac{\rho}{\epsilon}} {1-\frac{Ng_s}{2\pi} \,\log
\frac{\rho}{\epsilon}} \right)\,\theta  \label{csol}
\end{eqnarray}
where $g_s$ is the string coupling constant and $\epsilon$ is a
regulator. Notice that the explicit appearance of the angle $\theta$
in the above expressions of $C_0$ and $c$ implies that this
supergravity solution is not invariant under rotations in the
transverse plane, a fact that is similar to what has been recently
emphasized in Ref.~\cite{KOW} for the fractional D-branes of the
conifold.

Let us now consider the world-volume theory of the fractional D3/D7
brane system which, in the limit $\alpha' \to 0$, reduces to
${\cal{N}}=2$ super Yang-Mills theory with gauge group $SU(M)$ and $N$
fundamental hypermultiplets in four dimensions. In particular, the
action $S_{\rm YM}$ for the bosonic fields in the vector multiplet can
be simply obtained by taking the Dirac-Born-Infeld action plus the
Wess-Zumino term of a fractional D3 brane and expanding it in the
supergravity background previously considered. Taking the limit
$\alpha' \to 0$ and keeping fixed the combination
\begin{equation}
\phi = (2 \pi \alpha')^{-1}\,z \label{rel91}
\end{equation}
which plays the role of the scalar field of the ${\cal{N}}=2$ vector
multiplet, we easily find \cite{d3d7}
\begin{equation}
S_{\rm YM}= -\,\frac{1}{g^{2}} \int d^4 x  \left\{ \frac{1}{4}
 F_{\alpha \beta}^a F^{\alpha \beta}_a + \frac{1}{2} \partial_{\alpha}
 {\overline\phi}\, \partial^{\alpha} \phi \right\} + \frac{\theta_{\rm
 YM}}{32 \pi^2} \int d^4 x F_{\alpha \beta}^a {\widetilde{F}}^{\alpha
 \beta}_a \label{bound53}
\end{equation}
where \footnote{We take this opportunity to correct a sign misprint in
Eq.(5.5) of Ref.~\cite{d3d7}.}
\begin{eqnarray}
\frac{1}{g^{2}  } &=& \frac{{\rm e}^{-\varphi}\,b}{16\pi^3\alpha'g_s}
= \frac{1}{8\pi g_s} + \frac{2M-N}{8 \pi^2} \log \frac{\rho}{\epsilon}
\label{runn23} \\ \nonumber \\ \theta_{\rm YM} &=&
\frac{c+C_{(0)}\,b}{2\pi\alpha'g_s} =-(2M-N) \; \theta
~~.\label{runn24}
\end{eqnarray}
Inserting these expressions in Eq.(\ref{gau63}), we can see that
$\tau_{\rm YM}$ is a simple holomorphic function of $z$, namely
\begin{equation}
\tau_{\rm YM}  = {\rm i} \,\frac{2M-N}{2 \pi} \log \frac{z}{\rho_e}
 \label{gau42}
\end{equation}
where $\rho_e= \epsilon\, {\rm e}^{-\pi/(2M-N)g_s}$ is the distance in the
$z$-plane where the enhan{\c{c}}on phenomenon takes place
\cite{enhancon}. Finally, using Eqs.(\ref{tra34}) and (\ref{rel91}) we
deduce that the scale and chiral transformations are realized on the
supergravity coordinate $z$ as follows
\begin{equation}
z \rightarrow \mu\,{\rm e}^{2\,{\rm i} \,\alpha} \,z ~~,\label{tra32}
\end{equation}
and hence the field theory results (\ref{chisca3}) and (\ref{tra39})
are precisely reproduced from the supergravity classical solution.

The above analysis can be easily generalized by discussing  a more
general bound state in which both types of fractional D3 branes
present in the ${\bf{Z}_2}$ orbifold are considered \cite{REV}. The
most general configuration one can have is made of $N_1$ branes of
type 1 and $N_2$ branes of type 2. In this case the world-volume
theory is a ${\cal{N}}=2$ Yang-Mills theory with gauge group $SU(N_1)
\times SU(N_2 )$~\footnote{We neglect a diagonal $U(1)$ factor which
is decoupled, and the relative $U(1)$ which are subleading in the
large $N$-limit.} and two hypermultiplets in the bifundamental
representations $({\bf N_1},{\bf{\overline N}_2})$ and $({\bf
{\overline N}_1},{\bf N_2})$ respectively. In this case our previous
analysis can be generalized in a straightforward way and the correct
results are simply obtained by replacing in all formulas the quantity
$(2M-N)$ with $2(N_1-N_2)$ if we refer to the first factor of the
gauge group, or with $2(N_2-N_1)$ if instead we refer to the second
factor of the gauge group.

Finally, we would like to point out that the method of obtaining the
$\beta$-function and the chiral anomaly from the fractional brane
solutions as we have described it now, does not rely on the use of the
probe technique which instead was employed in Refs.~\cite{d3,d3d7,marco},
and in particular does not require that the analysis be made in the
Coulomb branch of the ${\cal{N}}=2$ theory. The only necessary ingredients
are the holographic identification of a world-volume field in terms of
bulk supergravity quantities, as we did for example in Eq.(\ref{rel91}),
and its behavior under scale and chiral transformations. Everything else
then follows from the explicit expressions Eqs.(\ref{runn23}) and
(\ref{runn24}) of the gauge coupling constant and the $\theta$-angle in
terms of the supergravity fields, expressions that are dictated by the
low-energy limit of the world-volume action of the fractional branes. Thus
this method can be in principle applied also to ${\cal{N}}=1$ models as we
will discuss in the next section.

%%%%%%%%%%%%%%%%%%%%%%%%%%%%%%%%%%%%%%%%%%%%%%%%%%%%%%%%%%%%%%%%%%%%%%
\section{${\cal{N}}=1$ gauge theories and fractional branes in ${\bf Z}_2
\times {\bf Z}_2$ orbifolds}

Let us now consider type IIB string theory in the orbifold ${\bf
C}^3/({\bf Z}_2\times {\bf Z}_2)$ which provides the simplest set up
to realize supersymmetric gauge theories with ${\cal N}=1$
supersymmetry in four dimensions by means of fractional D3 branes,
whose supergravity solution was obtained in Ref.~\cite{FERRO}.  For
definiteness we take the orbifold directions to be $x^4,...,x^9$,
introduce three complex coordinates defined by
\begin{equation}
z_1\equiv x^4+{\rm i}\,x^5=\rho_1\,{\rm e}^{{\rm i}\,\theta_1} ~~,~~
z_2\equiv x^6+{\rm i}\,x^7=\rho_2\,{\rm e}^{{\rm i}\,\theta_2}~~,~~
z_3\equiv x^8+{\rm i}\,x^9=\rho_3\,{\rm e}^{{\rm i}\,\theta_3} ~~,
\label{zetas}
\end{equation}
and consider fractional D3 branes that are completely transverse to
the orbifold, {\it i.e.} that are extended along $x^\alpha$ with
$\alpha=0,1,2,3$. As explained in Refs.~\cite{fiol,FERRO}, there are
four types of such fractional D3 branes corresponding to the four
irreducible representations of ${\bf Z}_2 \times {\bf Z}_2$, none of
which is free to move in the transverse space. The most general
configuration we can consider is therefore a stack of $N_1$ branes of
type 1, $N_2$ of type 2, $N_3$ of type 3 and $N_4$ of type 4, all
located at the orbifold fixed point. On the world volume of this bound
state there is a four dimensional ${\cal N}=1$ Yang-Mills theory with
gauge group $SU(N_1)\times SU(N_2)\times SU(N_3)\times
SU(N_4)$~\footnote{Again we do not consider irrelevant $U(1)$
factors.} and bifundamental matter. In particular for each factor
$SU(N_I)$ of the gauge group one finds 6 chiral multiplets, 3 of them
transforming in the bifundamental representation
($\bf{N_I},{\bf{\overline{N}_J}}$) and 3 in the conjugate
representation ($\bf{\overline{N}_I},{\bf{ N_J}}$) with $J\not=I$, for
a total of 12 chiral multiplets. In the explicit string realization
these fields are equipped with a suitable $4\times 4$ Chan-Paton
matrix that specifies on which of the four types of branes the two
end-points of the open string are attached. Taking this fact into
account and picking the same complex structure as in Eq.(\ref{zetas}),
under the same conventions as those of Ref.~\cite{FERRO}, one can
rearrange the 12 chiral multiplets into three $4\times 4$ matrices
given by
\begin{equation}
\Phi_1= \left( \begin{array}{cccc} \!\! 0\!\! & \!\!A_1\!\! & \!\!0
&\!\! 0\!\!  \\\!\! B_1 \!\!&\!\! 0 \!\!& \!\!0 \!\!&\!\! 0\!\!\\ \!\!
0 \!\!&\!\! 0 \!\!&\!\! 0 \!\!&\!\! C_1 \!\! \\ \!\!0 \!\!& \!\!0
\!\!&\!\! D_1\!\! & \!\!0 \!\!  \end{array} \right),~ \Phi_2= \left(
\begin{array}{cccc} \!\! 0\!\! & \!\!0\!\! & \!\!A_2 &\!\! 0\!\!
\\\!\! 0 \!\!&\!\! 0 \!\!& \!\!0 \!\!&\!\! B_2\!\!\\ \!\! C_2
\!\!&\!\! 0 \!\!&\!\! 0 \!\!&\!\! 0 \!\! \\ \!\!0 \!\!& \!\!D_2
\!\!&\!\! 0\!\! & \!\!0 \!\!  \end{array} \right),~ \Phi_3= \left(
\begin{array}{cccc} \!\! 0\!\! & \!\!0\!\! & \!\!0 &\!\! A_3\!\!
\\\!\! 0 \!\!&\!\! 0 \!\!& \!\!B_3 \!\!&\!\! 0\!\!\\ \!\! 0 \!\!&\!\!
C_3 \!\!&\!\! 0 \!\!&\!\! 0 \!\! \\ \!\!D_3\!\!& \!\!0 \!\!&\!\! 0\!\!
& \!\!0 \!\!  \end{array} \right)
\label{sc12}
\end{equation}
where $A_i,\cdots,D_i$ are each a chiral multiplet. The position of
these multiplets inside the matrices indicates from which types of
open strings they originate, for example $A_1$ arises from strings
stretched between branes of type 1 and branes of type 2, whereas $C_3$
from strings stretched between branes of type 3 and branes of type 2
and so on. Each field matrix $\Phi_i$ encodes those chiral
superfields having dynamics in the $z_i$ plane.

There are several important points that we would like to
emphasize. First of all, the Lagrangian of this ${\cal N}=1$ theory
contains a {\it cubic} superpotential of the form $W={\rm Tr}(\Phi_1
[\Phi_2,\Phi_3])$, which is renormalizable in the UV.  This is to be
contrasted to what happens in the conifold theory where the matter
fields have a {\it quartic} unrenormalizable superpotential
\cite{KW,KS}. Secondly, the Lagrangian of the orbifold theory is
classically invariant under scale transformations of the energy and
$U(1)_R$ transformations. Due to the presence of the cubic
superpotential, the superfields $\Phi_i$ have $R$-charge 2/3, and
hence their scalar components $\phi_i$ have charge 2/3 while the
chiral fermions have charge $-1/3$. Therefore, under a scale
transformation with parameter $\mu$ and a $U(1)_R$ transformation with
parameter $\alpha$ we have
\begin{equation}
\phi_i \rightarrow \mu\,{\rm e}^{\frac{2}{3}\,{\rm i}\, \alpha}
\,\phi_i \label{tra341}
\end{equation}
for $i=1,2,3$. As is well known, these transformations become
anomalous in the quantum theory.  Indeed, focusing for simplicity on
the first factor of the gauge group (similar considerations hold  for
any gauge groups), we find a scale anomaly proportional to the
(Wilsonian) $\beta$-function
\begin{equation}
\beta(g) = - \frac{(3N_1-N_2-N_3-N_4)}{16\pi^2}\,g^3
\label{beta1}
\end{equation}
where $g$ is the running coupling constant of $SU(N_1)$, and a
$U(1)_R$ anomaly given by
\begin{equation}
\partial_{\alpha} J^{\alpha}_{R} =2 \left( N_1 - \frac{1}{3} \left(
N_2 + N_3 + N_4 \right)\right)\,\, q(x) 
\label{ano781}
\end{equation}
where $q(x)$ is the topological charge density for $SU(N_1)$ (see
Eq.(\ref{ano78})). Just like in the ${\cal N}=2$ theories, also in
this case we can combine the effect of the  scale and chiral anomalies
together by writing
\begin{equation}
\tau_{\rm YM}  \rightarrow \tau_{\rm YM} + {\rm i}
\,\frac{(3N_1-N_2-N_3-N_4)}{2\pi} \left(\log\mu+\frac{2}{3}{\rm
i}\,\alpha \right)
\label{chisca31}
\end{equation}
where $\tau_{\rm YM}$ is defined as in Eq.(\ref{gau63}) in terms of
the coupling constant and $\theta$-angle of the first factor of the
gauge group. Of course similar expressions hold for the other factors
and can be obtained from the previous formulas in a straightforward
way. Notice that while the chiral anomaly is a one-loop effect, in
${\cal N}=1$ gauge theories the $\beta$-function receives corrections
at all loops. The reason why it is possible to construct the complex
combination (\ref{chisca31}) is because here we are discussing the
Wilsonian $\beta$-function which is perturbatively exact at one loop~\cite{SV}.

Let us now consider the supergravity background corresponding to our
bound state of fractional D3 branes~\cite{FERRO}. This is
characterized by a metric of the form
\begin{equation}
ds^2 = H^{-1/2}\, \eta_{\alpha\beta}\,d x^\alpha dx^\beta + H^{1/2}
\,\delta_{\ell m} dx^\ell dx^m  ~~,
\label{metric1}
\end{equation}
a R-R 4-form $C_{(4)}$ and three pairs of scalars $b_i$ and $c_i$
($i=1,2,3$) which correspond to the components of the 2-forms
$B_{(2)}$ and $C_{(2)}$ along the anti-self dual forms
$\omega_{(2)}^i$ associated to the three exceptional vanishing cycles
of the orbifold, namely
\begin{equation}
C_{(2)} = c_i~ \omega^i_{(2)}~~~~~,~~~~B_{(2)} = b_i~ \omega^i_{(2)}
\label{wra981}
\end{equation}
The explicit form of the solution can be found in Ref.~\cite{FERRO};
here we simply recall that
\begin{eqnarray}
b_i&=& (2 \pi^2{\alpha'})\; \left(1+\frac{2g_s}{\pi}\,f_i(N_I) \, \log
\frac{\rho_i}{\epsilon}\right)\label{bsol1}\\ c_i&=&-
(4\pi\alpha')\,g_s\,f_i(N_I)\,\theta_i \label{csol1}
\end{eqnarray}
where
\begin{eqnarray}
f_1(N_I)&=& N_1+N_2-N_3-N_4~~,\nonumber\\f_2(N_I)&=&
N_1-N_2+N_3-N_4~~,\label{fi}\\ f_3(N_I)&=& N_1-N_2-N_3+N_4~~.\nonumber
\end{eqnarray}
As we mentioned before, the world-volume action of our bound state of
fractional D3 branes in the limit $\alpha'\to 0$ reduces to a ${\cal
N}=1$ super Yang-Mills theory in four dimensions with gauge group
$SU(N_1)\times SU(N_2)\times SU(N_3)\times SU(N_4)$ and bifundamental
matter.  The bosonic part of this action can be obtained by expanding
the Dirac-Born-Infeld action plus the Wess-Zumino term for fractional
D3 branes in the corresponding supergravity background, and then
observing that the complex coordinates $z_i$ of Eq. (\ref{zetas})  can
be traded for the scalar components $\phi_i$ of the chiral superfields
$\Phi_i$ of Eq. (\ref{sc12}), similarly to what we have done in
Eq. (\ref{rel91}) with the scalar component of the ${\cal N}=2$ vector
multiplet. This correspondence can also be understood by  looking at
the explicit open-string realization of the scalars  $\phi_i$, each of
which is indeed related to a position in the $z_i$ plane
\cite{FERRO}. Using this identification between $\phi_i$ and $z_i$,
and Eq. (\ref{tra341}), we can find the realization of  the scale and
chiral transformations on the supergravity coordinates, namely
\begin{equation}
z_i \rightarrow \mu\,{\rm e}^{\frac{2}{3}\,{\rm i}\,\alpha}\,z_i
\label{tra}
\end{equation}
We now focus for simplicity on those terms of world-volume action
that, in the low-energy limit, depend only on the gauge fields of the
first factor $SU(N_1)$ of the gauge group. These terms are simply
\cite{FERRO}
\begin{equation}
S_{\rm YM}= -\,\frac{1}{4 g^{2}} \int d^4 x  \,F_{\alpha \beta}^a
 F^{\alpha \beta}_a +  \frac{\theta_{\rm YM}}{32 \pi^2} \int d^4 x
 \,F_{\alpha \beta}^a {\widetilde{F}}^{\alpha \beta}_a \label{bound531}
\end{equation}
where
\begin{eqnarray}
\frac{1}{g^2}&=& \frac{1}{8\pi
g_s}\left(\frac{1}{4\pi^2\alpha'}\sum_{i}b_i-1\right) = \frac{1}{16\pi
g_s}+\frac{1}{8\pi^2}\,\sum_{i}
f_i(N_I)\log\frac{\rho_i}{\epsilon}\nonumber \\ \theta_{\rm YM} &=&
\frac{1}{4\pi\alpha'g_s}\sum_{i}c_i=-\sum_i f_i(N_I)\,\theta_i
\label{par1}
\end{eqnarray}
Using these formulas, we see that the supergravity realization of  the
complex coupling $\tau_{\rm YM}$ is
\begin{equation}
\tau_{\rm YM} = {\rm i}\left[
\frac{1}{4g_s}+\frac{1}{2\pi}\sum_if_i(N_I)\log\frac{z_i}{\epsilon}\right]
\label{tau11}
\end{equation}
From this equation it is now immediate  to see that the field theory
result (\ref{chisca31})  is correctly reproduced if we use the
transformation (\ref{tra})  and the explicit definitions of the
functions $f_i(N_I)$ given in Eq. (\ref{fi}).

We conclude by observing that a pure ${\cal N}=1$ super Yang-Mills
theory in four dimensions can be realized on  a stack of $M$
fractional D3 branes of just one type, for example by putting $N_1=M$
and $N_2=N_3=N_4=0$ in our previous analysis. Therefore, the method we
have described allows to obtain the correct $\beta$-function and
chiral anomaly  of this theory from the supergravity solution, namely
\begin{equation}
\frac{1}{g^{2}} \rightarrow \frac{1}{g^{2}}
+\frac{3M}{8\pi^2}\,\log\mu~~~~~{\rm and}~~~~~\\ \theta_{\rm YM}
\rightarrow \theta_{\rm YM} -2M\, \alpha~~. \label{tra391}
\end{equation}
Notice that the equation above correctly accounts for the breaking of
the $U(1)$ R-symmetry to ${\bf Z}_{2M}$. To our knowledge this is the
first quantitative derivation of these results for pure ${\cal N}=1$
super Yang-Mills theory using a supergravity dual background.  Since
these are UV results, the naked singularity of the supergravity
solution  does not play any role. It would be very interesting to
explore the possibility of resolving this singularity, for example by
suitably  deforming the orbifold ${\bf C}^3/({\bf Z}_2\times {\bf
Z}_2)$,  and see whether in this way one can get some information on
the IR behavior of the dual gauge theory.

%%%%%%%%%%%%%%%%%%%%%%%%%%%%%%%%%%%%%%%%%%%%%%%%%%%%%%%%%%%%%%%%%%%%%%%%%%%%
\vskip 0.8cm
\noindent
{\large {\bf Acknowledgments}}
\vskip 0.2cm
\noindent
M.F. and A.L. thank M. Bill\'o and I. Pesando and M.B. thanks
G. Ferretti, E. Imeroni, E. Lozano-Tellechea and J.L. Petersen for useful
discussions and exchange of ideas. M.B. is supported by  a EC Marie
Curie Postdoc Fellowship under contract number HPMF-CT-2000-00847.

%%%%%%%%%%%%%%%%%%%%%%%%%%%%%%%%%%%%%%%%%%%%%%%%%%%%%%%%%%%%%%%%%%%%%%


\begin{thebibliography}{99}

\bibitem{KN}  I.~R.~Klebanov and N.~A.~Nekrasov,
\emph{Gravity duals of fractional branes and logarithmic RG flow},
Nucl.\ Phys. {\bf B574} (2000) 263, {\tt hep-th/9911096}.
%%CITATION = HEP-TH 9911096;%%
%
\bibitem{KT} I.R. Klebanov and A.A. Tseytlin, \emph{Gravity Duals of 
Supersymmetric SU(N) X SU(N+M) Gauge Theories}, Nucl. Phys. {\bf B578} 
(2000) 123, {\tt hep-th/0002159}.
%%CITATION = HEP-TH 0002159;%%
%
\bibitem{KS} I.R. Klebanov and M.J. Strassler, \emph{Supergravity and a 
Confining Gauge Theory: Duality Cascades and $\chi$SB-Resolution of 
Naked Singularities}, JHEP {\bf 0008} (2000) 052, {\tt hep-th/0007191}.
%%CITATION = HEP-TH 0007191;%%
%
\bibitem{KW} I.~R.~Klebanov and E.~Witten, \emph{Superconformal 
field theory on threebranes at a Calabi-Yau  singularity}, Nucl.Phys. 
{\bf B536} (1998) 199, {\tt hep-th/9807080}.
%%CITATION = HEP-TH 9807080;%%
%
\bibitem{HKO} C.P. Herzog, I.R. Klebanov and P. Ouyang, 
\emph{Remarks on the warped deformed conifold}, {\tt hep-th/0108101}.
%%CITATION = HEP-TH 0108101;%%"
%
\bibitem{KOW}
I.R. Klebanov, P. Ouyang and E. Witten, 
\emph{A Gravity Dual of the Chiral Anomaly}, {\tt hep-th/0202056}.
%%CITATION = HEP-TH 0202056;%%
%
\bibitem{d3}
M.~Bertolini, P.~Di~Vecchia, M.~Frau, A.~Lerda, R.~Marotta and
I.~Pesando, \emph{Fractional D-branes and their gauge duals}, JHEP
{\bf 02} (2001) 014, {\tt hep-th/0011077}.
%%CITATION = HEP-TH 0011077;%%
%
\bibitem{polch}
J.~Polchinski, \emph{N = 2 gauge-gravity duals}, Int. J. Mod.
Phys. {\bf A16} (2001) 707, {\tt hep-th/0011193}.
%%CITATION = HEP-TH 0011193;%%
%
\bibitem{grana2}
M. Gra\~na and J. Polchinski,
{\em Gauge/gravity duals with holomorphic dilaton}, {\tt
hep-th/
0106014}.
%%CITATION = HEP-TH 0106014;%%
%
\bibitem{d3d7}
M. Bertolini, P. Di Vecchia, M. Frau, A. Lerda and R. Marotta, 
\emph{N=2 gauge theories on systems of fractional D3/D7 branes}, 
Nucl. Phys. {\bf B621} (2002) 157, {\tt hep-th/0107057}.
%%CITATION = HEP-TH 0107057;%%
%
\bibitem{marco} M. Bill\`o, L. Gallot and A. Liccardo, {\em Classical 
geometry and gauge duals for fractional branes on ALE spaces}, 
Nucl. Phys. {\bf B614} (2001) 254,
{\tt hep-th/0105258}.
%%CITATION = HEP-TH 0105258;%%
%
\bibitem{REV} M. Bertolini, P. Di Vecchia and R. Marotta, 
\emph{N=2 four-dimensional gauge theories from fractional branes}, 
{\tt hep-th/0112195}.
%
%%CITATION = HEP-TH 0012035;%%
\bibitem{FERRO}
M. Bertolini, P. Di Vecchia, G. Ferretti and R. Marotta, 
\emph{Fractional Branes and $N=1$ Gauge Theories}, {\tt hep-th/0112187}.
%
\bibitem{HOUCHES98}
P. Di Vecchia, 
\emph{Duality in supersymmetric $N=2,4$ gauge theories}, {\tt hep-th/9803026}.
%
\bibitem{PN} P. Di Vecchia, R. Musto, F. Nicodemi and R. Pettorino, 
\emph{The anomaly term in the N=2 supersymmetric gauge theory}, Nucl. Phys. 
{\bf B252} (1985) 635.
%
\bibitem{enhancon}
C.V. Johnson, A.W. Peet and J. Polchinski, \emph{Gauge theory and
the excision of repulson singularities}, Phys. Rev. {\bf D61}
(2000) 086001, {\tt hep-th/9911161}.
%%CITATION = HEP-TH 9911161;%%
%
\bibitem{fiol}  David R. Morrison and M. Ronen Plesser,
\emph{ Non-Spherical Horizons, I}, Adv. Theor. Math. Phys. 
{\bf 3} (1999) 1, {\tt hep-th/ 9810201}.
%
%%\cite{Shifman:1986zi}
\bibitem{SV} M.~A.~Shifman and A.~I.~Vainshtein,
\emph{Solution Of The Anomaly Puzzle In Susy Gauge Theories And The 
Wilson Operator Expansion}, Nucl.Phys. {\bf B277} (1986) 456 
[Sov.Phys. JETP {\bf 64} (1986) 428].
%%CITATION = NUPHA,B277,456;%%

\end{thebibliography}
\end{document}